\documentclass[aps,showpacs,floatfix, nofootinbib]{revtex4-1}
\usepackage{amsfonts}
\usepackage{amsmath}
\usepackage{amssymb}
\usepackage{graphicx}
\usepackage{color}

\newcommand{\Ex}[1]{(\ref{#1})}
\newcommand{\beq}{\begin{equation}}
\newcommand{\eeq}{\end{equation}}
\newcommand{\beqa}{\begin{eqnarray}}
\newcommand{\eeqa}{\end{eqnarray}}

\begin{document}

\title{A microscopic approach to nonlinear Reaction-Diffusion: \\
the case of morphogen gradient formation}
\author{Jean Pierre Boon}
\email{jpboon@ulb.ac.be}
\homepage{http://poseidon.ulb.ac.be}
\author{James F. Lutsko}
\email{jlutsko@ulb.ac.be}
\homepage{http://www.lutsko.com}
\affiliation{Center for Nonlinear Phenomena and Complex Systems CP 231\\
Universit\'e Libre de Bruxelles, 1050 - Bruxelles, Belgium}
\author{Christopher Lutsko}
\affiliation{International School of Brussels, Kattenberg 19, 1170 Bruxelles, Belgium}

\begin{abstract}
We develop a microscopic theory for reaction-difusion (R-D) processes based
on a generalization of Einstein's master equation with a reactive term and
we show how the mean field formulation leads to a generalized R-D equation
with non-classical solutions. For the $n$-th order annihilation reaction $%
A+A+A+...+A\rightarrow 0$, we obtain a nonlinear reaction-diffusion equation
for which we discuss scaling and non-scaling formulations. We find steady
states with either solutions exhibiting 
long range power law behavior 
showing the relative dominance of sub-diffusion over reaction effects in
constrained systems, or conversely solutions 
with finite support of the concentration distribution describing situations
where diffusion is slow and extinction is fast. Theoretical results are
compared with experimental data for morphogen gradient formation.
\end{abstract}

\date{\today}
\maketitle


\section{Introduction}

{The random walk is the classical paradigm for the microscopic mechanism
underlying diffusive processes as demonstrated in 1905 by Einstein who
showed how the diffusion equation follows from the mean field formulation of
the microscopic random walk. Here we generalize the formulation for
situations where the diffusing particles are also subjected to a reactive
process. From the phenomenological viewpoint, when diffusion and reaction
are coupled, these processes are described by reaction-diffusion (R-D)
equations. For instance, the evanescence process ($A \rightarrow 0$) of
suspended particles diffusing in a non-reactive medium the concentration of
species $A$, $c(r;t)$, is described by the classical R-D equation 
\begin{equation}  \label{GRDan2}
\frac{\partial }{\partial t}c\left( r; t \right) = D\,\frac{\partial ^2}{%
\partial r^2} c\left( r; t \right) \,-\, k\,c\left( r; t \right) \,,
\end{equation}
where $D$ denotes the diffusion coefficient and $k$ the reaction
(evanescence) rate. This classical equation yields a steady state solution
showing spatial exponential decay of the concentration (but one can equally
consider the distribution function): $c \left( r\right)\,= \, c(0) \,\exp
\left({-\sqrt{{k}/{D}}}\,|r|\right)$ when particles are injected with a
constant flux at $r=0$. }

However there are many systems observed in nature where it seems logical to
use the language of reaction-diffusion, but where non-classical
distributions are found, i.e. the steady state spatial distributions are
non-exponential e.g. when the particles encounter obstacles or are retarded
in their diffusive motion, or because the reactive process is hindered or
enhanced by concentration effects. Such situations are ubiquitous in
chemical, rheological, biological,~... systems - a typical example being the
diffusion and degradation of a morphogen in cells during the early
developing stage {\cite{wartlick}} - and are certainly as commonly observed
as those that can be described by the idealized R-D system of Eq.(\ref%
{GRDan2}). This is why approaches to a more general description of R-D
phenomena have been proposed, and recent developments in this direction {%
\cite{abad, yuste}} are based (i) on a generalization of the diffusive
mechanism accounting for time delay effects or obstacles hindrance using the
continuous time random walk (CTRW) model and corresponding to a fractional
Fokker-Planck equation (FFP) or the fractional Brownian motion (FBM), and
(ii) on a space and time dependence of the reaction rate ($k \rightarrow
k(r;t)$). However the resulting expressions for the steady state
distribution have so far been subject to controversial comments expressing
that "CTRW theory is compatible with available experiment" {\cite{CTRW}} and
"that fractional Brownian motion is the underlying process" {\cite{FBM}} or
that "experimental results cannot be explained by a continuous time random
walk" {\cite{NotCTRW}} and "exclude fractional Brownian motion as a valid
description" {\cite{NotFBM}}. So the present state of the art certainly
appears somewhat confusing while it seems nevertheless clear that a general
R-D theory requires a generalization for both diffusion and reaction.

Here we present an alternative approach by developing a microscopic theory
generalizing Einstein's master equation with a reactive term and we show how
the mean field formulation leads to the nonlinear R-D equation with
non-classical solutions. 
For the $n$-th order annihilation reaction $A + A + A + ... + A \rightarrow
0 $, we obtain the nonlinear reaction-diffusion equation (with no drift) 
\begin{equation}
\frac{\partial }{\partial t}c\left( r; t\right) = \frac{\partial}{\partial{r}%
}\,D\,\frac{\partial}{\partial{r}} c^{\alpha}\left( r; t\right)- k \,
c^{n}\left( r; t\right) \,,
\end{equation}%
for which we discuss scaling and non-scaling formulations and the
corresponding range of values of the nonlinear exponents. We obtain steady
state solutions of the form $c(r) = c(0) \left( 1 + C_{\alpha, n}(D,k)\,{r }/%
{\nu} \right)^{-\nu}$ where $\nu = \frac{2}{n - \alpha}$, giving long range
power law behavior (for $n>\alpha$) showing the relative dominance of
sub-diffusion over reaction effects in constrained systems, or conversely
(for $n<\alpha<n+1$) leading to finite support of the concentration $c(r)$
describing the situation where diffusion is slow and extinction is fast. 
An experimental example of morphogen gradient formation is discussed.

\section{Generalized master equation}

We consider a diffusive process where particles are subject to annihilation
using the microscopic approach of Einstein's original random walk model. For
simplicity we consider a one-dimensional lattice where the particle hops to
the nearest neighboring site (left or right) in one time step, and can then
also be annihilated by a some reactive process as described by the discrete
equation 
\begin{equation}
n^{\ast }(r;t+1)=\xi _{-}\,n^{\ast }(r+1;t_{\;-})+\xi _{+}\,n^{\ast
}(r-1;t_{\;-})-\xi _{R}\,n^{\ast }(r;t_{\;+})\,,  \label{RW_Eq}
\end{equation}%
where the Boolean variable $n^{\ast }(r;t)=\{0,1\}$ denotes the occupation
at time $t$ of the site located at position $r$ and $\xi _{\pm }$ is a
Boolean random variable controlling the particle jump between neighboring
sites ($\xi _{+}+\,\xi _{-}\leq 1$), while $\xi _{R}$ is the reactive
Boolean operator controling particle annihilation. 
The mean field description follows by ensemble averaging Eq.(\ref{RW_Eq})
with $\langle n^{\ast }(r;t)\rangle =n(r;t)$, $\langle \xi _{\pm }\rangle
=P_{i}$, and $\langle \xi _{R}\rangle =R_{i}$, where $i$ is an index for the
position; using statistical independence of the $\xi $'s and $n^{\ast }$,
and extending the possible jump steps over the whole lattice, we obtain 
\begin{equation}
n(r;t+\delta t)\,=\,\sum_{j=-\infty }^{+\infty }P_{j}(r-j\delta
r;t)n(r-j\delta r;t)\,-\,R\,(r;t)\,n(r;t)\;,  \label{Master_Eq0}
\end{equation}%
where $P_{j}(r-j\delta r)$ denotes the probability of a jump of $j$ sites
from site $r-j\delta r$, and $R\,(r)$ the annihilation probability at site $%
r $; the number density is $n\left( r;t\right) $ so that $n\left( r;t\right)
dr $ is the expected number of particles to find in the interval $\left[
r-dr/2,r+dr/2\right] $. Note that in a closed system, i.e. without the
second term on the right, the total number of particles, $N$, is constant so
that one can divide through by this number to express the master equation in
terms of $f(r,t)=n(r,t)/N$ , the probability density. Alternatively, if the
system contains multiple components, then a more useful concept is the
concentration. For example, if there are two components, one of which is the
solvent and the other the solute, then the solute concentration would be $%
c(r,t)=n(r,t)/(n(r,t)+n_{s}(r,t))$, where $n_{s}(r,t)$ is the local number
density for the solute. In the common case that the solvent is uniform and
stationary, $n_{s}(r,t)=n_{s}$, and that the solute is relatively dilute, $%
n_{s}>>n(r,t)$, one has, to first approximation, $c(r,t)=n(r,t)/n_{s}$,
which is what we will use in the following.

In the classical case, the jump probabilities are constants, $%
P_{j}(r-j\delta r;t)=p_{j}\geq 0$ with $\sum_{j=-\infty }^{\infty }p_{j}=1$,
as is the reaction probability, $R\,(r;t)=p_{R}$ with $1\geq p_{R}\geq 0$.
We take into account the configurational complexity of the reactive medium
by allowing for the possibility that both the jump probabilities and the
reaction probability are modified by interaction between the particles. This
is modeled by writing ~$P_{j}\left( r-j\delta r;t\right) =p_{j}\,F\left[
c\left( r-j\delta r;t\right) \right] $, with $j\neq 0$ and ~$R\,\left(
r;t\right) =p_{R}\,G\left[ c\left( r;t\right) \right] $ 
giving the Generalized Master Equation 
\begin{equation}
c\left( r;t+\delta t\right) \,-c\left( r;t\right) =\sum_{j=-\infty
}^{+\infty }\,\left( p_{j}F\left[ c\left( r-j\delta r;t\right) \right]
\left( c\left( r-j\delta r;t\right) \,-c\left( r;t\right) \right) \right)
\,-\,p_{R}\,G\left[ c\left( r;t\right) \right] \,c(r;t)\,.  \label{Master_Eq}
\end{equation}%
Notice that in order to retain their nature as probabilities, the functions $%
F\left[ c\right] $ and $G\left[ c\right] $ must both be greater than zero
and less than one for all values of their arguments.

\section{Diffusion and reaction}

\subsection{Generalized diffusion equation}

Considering the diffusive process alone, it was shown {\cite{lutsko-boon}
that the generalized diffusion equation that follows from Eq.(\ref{Master_Eq}%
) (without the second term on the \textit{r.h.s}) is 
\begin{align}
\frac{\partial c}{\partial t}+C\frac{\partial }{\partial r}\left( xF\left(
x,x\right) \right) _{c}& =D\frac{\partial }{\partial r}\left( \frac{\partial
xF\left( x,y\right) }{\partial x}-\frac{\partial xF\left( x,y\right) }{%
\partial y}\right) _{c}\frac{\partial c}{\partial r}  \notag \\
& +\frac{C^{2}\delta t}{2}\frac{\partial }{\partial r}\left( \frac{\partial
xF\left( x,y\right) }{\partial x}-\frac{\partial xF\left( x,y\right) }{%
\partial y}-\left( \frac{\partial xF\left( x,x\right) }{\partial x}\right)
^{2}\right) _{c}\frac{\partial c}{\partial r}\,,  \label{diff_eq}
\end{align}%
with the compact notation $\left( ...\right) _{c}=\left( ...\right)
_{x=c\left( r,t\right) ,y=c\left( r,t\right) }$. 
Here $C=\left( \sum_{j}\,j\,p_{j}\,\right) \,\frac{\delta r}{\delta t}$ is
the advection speed and $D~=~\left( \sum_{j}j^{2}p_{j}\,\right) \,\frac{%
\left( \delta r\right) ^{2}}{2\delta t}$ is the diffusion coefficient. In {%
\cite{lutsko-boon}} it was also shown that 
the existence of a scaling solution $c\left( r;t\right) =t^{-\gamma /2}\phi
\left( r/t^{{\gamma /2}}\right) $ demands that $F[c]\sim c^{\eta }$ in which
case the scaling exponent is 
$\gamma =\frac{2}{2+\eta }$; since the jump probabilities $P_{j}=p_{j}F\left[
c\right] $ must be $\leq 1$, one must have $\eta >0$, that is $\gamma <1$,
which is the signature of sub-diffusion \footnote{For $\eta =0$, one has $\gamma =1$, i.e.
classical diffusion. The case of super-diffusion will be presented elsewhere.}. 
We now combine the description of sub-diffusion (with no drift, i.e. $C=0$
in (\ref{diff_eq})) with reactive processes. }

\subsection{Scaling reaction-diffusion}

Starting from the generalized master equation (\ref{Master_Eq}), we proceed
along the lines of derivation of the generalized diffusion equation given in 
{\cite{lutsko-boon}}. Performing a multiple scale expansion up to second
order, we obtain the general form of the reaction-diffusion (R-D) equation
(with no drift and with reaction rate $k=p_{R}\,\frac{1}{\delta t}$): 
\begin{equation}
\frac{\partial }{\partial t}c\left( r;t\right) =D\frac{\partial ^{2}}{%
\partial {r^{2}}}\,\left( F[c\left( r;t\right) ]c\left( r;t\right) \right)
-\,k\,G[c\left( r;t\right) ]\,c\left( r;t\right) \,.  \label{GRD}
\end{equation}%
%
%
%
%
%
%

As for the generalized diffusion equation {\cite{lutsko-boon}}, we ask under
which conditions there is a scaling solution to equation (\ref{GRD}) of the
form $c\left( r;t\right) =t^{-\gamma /2}\phi \left( r/t^{{\gamma /2}}\right)
=t^{-{\gamma /2}}\phi \left( x\right) $. Expressing the time and space
derivatives in terms of $x$, Eq.(\ref{GRD}) can be written as 
\begin{equation}
-\gamma \frac{d}{dx}x\phi \left( x\right) =2\,Dt^{1-\gamma }\frac{d^{2}}{%
dx^{2}}F\left( t^{-{\gamma /2}}\phi \left( x\right) \right) \phi \left(
x\right) -\,k\,t\,G\left( t^{-{\gamma /2}}\phi \left( x\right) \right) \phi
\left( x\right) \,.  \label{x_D_Eq}
\end{equation}%
The time-dependence on the right can only be eliminated if $F(c)$ and $G(c)$
have a functional power law form: $F(c)=c^{\alpha -1}=t^{(1-\alpha ){\gamma
/2}}\,\phi ^{\alpha -1}$ and $G(c)=c^{n-1}=t^{(1-n){\gamma /2}}\,\phi ^{n-1}$%
, for some numbers $\alpha \geq 1$ and $n\geq 1$ ; hence we must have $%
1=t^{1-\gamma }t^{(1-\alpha )\,{\gamma /2}}$, and $1=t\;t^{(1-n)\,{\gamma /2}%
}$, that is 
\begin{equation}
\gamma =\frac{2}{\alpha +1}\;;\;\;\;\;n-1=\frac{2}{\gamma }\;.  \label{gamma}
\end{equation}%
Thus, according to scaling consistency the exponents should be such that $%
n=\alpha +2$. When $\alpha > 1$, we have anomalous diffusion: $%
\left\langle r^{2}\right\rangle \sim t^{\frac{2}{\alpha +1}}$ (and more
generally $\left\langle r^{m}\right\rangle \sim t^{\frac{m}{\alpha +1}}$),
and the reaction term goes like $\sim - k\,\phi ^{n}$. More explicitly,
using in (\ref{x_D_Eq}) the reduced variable 
\begin{equation}
\zeta =xk^{-\gamma /2}\,\sqrt{\frac{\,k}{\,D}}\;,  \label{reduced}
\end{equation}%
we obtain the scaled equation 
\begin{equation}
\frac{d^{2}}{d\zeta ^{2}}\,\phi ^{\alpha }(\zeta )\,+\,\frac{1}{\alpha +1}%
\frac{d}{d\zeta }(\zeta \,\phi (\zeta ))\,-\,\phi ^{n}(\zeta )=0\,,
\label{scaled_Eq}
\end{equation}%
which can be rewritten in terms of the original variables ($r$ and $t$) to
give 
\begin{equation}
\frac{\partial }{\partial t}c\left( r;t\right) =\frac{\partial }{\partial {r}%
}\,\,D\,\frac{\partial }{\partial {r}}c^{\alpha }\left( r;t\right)
-\,kc^{n}\left( r;t\right) \,.  \label{GRD0}
\end{equation}%
Without the reactive term, i.e. with $k=0$, this reduces to our previous
generalized diffusion equation in the absence of drift. Equation(\ref{GRD0})
is the generalized reaction-diffusion equation.

\section{Steady-state distributions}

In this Section we explore Eq.(\ref{GRD0}) as a natural extension of our
previous description of generalized diffusion to include extinction. Because
we are not solely interested in scaling solutions in this case, we will
allow for arbitrary exponents $\alpha \geq 1$ and $n > 0$.

\subsection{Boundary conditions}

One frequently studied problem is that of a semi-infinite sytem with
constant injection of particles at the boundary. To be specific, we use the
interval $[0,\infty ]$ and note that  the rate of change of the total
number of particles in is simply %
\begin{eqnarray}
\frac{dN\left( t\right) }{dt} &=&n_{s}\int_{0}^{\infty }\,\frac{\partial
c\left( r;t\right) }{\partial t}dr \\
&=&Dn_{s}\left. \,\frac{\partial c^{\alpha }\left( r;t\right) }{\partial {r}}%
\right\vert _{r\rightarrow \infty }-Dn_{s}\left. \frac{\partial c^{\alpha
}\left( r;t\right) }{\partial {r}}\right\vert _{r=0}-kn_{s}\int_{0}^{\infty
}\,\frac{\partial c^{n}\left( r;t\right) }{\partial t}dr \,.  \notag
\end{eqnarray}%
The first term on the right is the rate at which matter leaves the system
via the boundary at infinity:\ we will assume that the concentration goes to
zero sufficiently fast at infinity so that this term is zero - an assumption
that will have to be checked a posteriori. The second term on the right is
the rate at which particles are injected at the left boundary and the last
term is the rate at which particles are removed by the extinction process.
Our boundary condition will be to control the rate at which particles are
injected so we set 
\begin{equation}
\left( \frac{dN\left( t\right) }{dt}\right) _{in}\equiv j_{0}=-Dn_{s}\left. 
\frac{\partial c^{\alpha }\left( r;t\right) }{\partial {r}}\right\vert _{r=0}
\end{equation}%
as the boundary condition of interest.

\subsection{Steady-state solution}

\label{ss_sol}

We now seek a steady state solution with this boundary condition, 
\begin{equation}
0=D\frac{\partial ^{2}}{\partial {r}^{2}}c^{\alpha }\left( r\right)
\,-\,kc^{n}\left( r\right) \;\;\;\;\;\;\mbox{with}\;\;\;\;\;\,-\,Dn_{s}%
\left. {\frac{\partial {c^{\alpha }}\left( r\right) }{\partial {r}}}%
\right\vert _{r=0}\,=\;j_{0}.
\label{ss_eq}
\end{equation}%
It is convenient to rewrite the problem with the change of variables 
\begin{equation}
r\rightarrow z\,=\,\sqrt{\frac{\,k}{D}}\,r\;\;\;;\;\;\;\;{j}_{0}\rightarrow {%
j}_{0}^{\ast }\,=\,\frac{{j}_{0}}{n_{s}\sqrt{kD}}\,,\;\;\;;\;\;\;\;c%
\rightarrow g=c^{\alpha }
\end{equation}%
so that the steady state equation has the simple form 
\begin{equation}
\frac{\partial ^{2}}{\partial z^{2}}g\left( z\right) \,=g^{\frac{n}{\alpha }%
}\left( z\right) \,\;\;\;\;\;\;\mbox{with}\;\;\;\;\;\left. {\frac{\partial {g%
}\left( z\right) }{\partial {z}}}\right\vert _{z=0}=-{j}_{0}^{\ast }.
\end{equation}%
This is integrated to get 
\begin{equation}
\frac{dg\left( z\right) }{dz}=\pm \sqrt{A+\frac{2\,\alpha }{\alpha +n}g^{%
\frac{\alpha +n}{\alpha }}\left( z\right) }\,.  \label{dgdz}
\end{equation}%
Recall that we assumed that the flux at infinity goes to zero. This means
that either $A=0$ and $\lim_{z\rightarrow \infty }g\left( z\right)=0 $ or
that $A<0$ and $\lim_{z\rightarrow \infty }g\left( z\right) $ is finite . We
rule out the latter case on the ground that without extinction we should get
purely diffusive behavior and that adding extinction should not cause an
increase in particles far from the source.

A second integration then gives the implicit solution 
\begin{equation}
\pm z\,=\int_{0}^{z}\frac{dg}{\sqrt{\frac{2\alpha }{\alpha +n}\,g^{\frac{%
\alpha +n}{\alpha }}}}=\sqrt{\frac{\alpha +n}{2\alpha }}\frac{2\alpha }{%
\alpha -n}\left( g^{\frac{\alpha -n}{2\alpha }}\left( z\right) -g^{\frac{%
\alpha -n}{2\alpha }}\left( 0\right) \right) \,,  \label{dgdz1}
\end{equation}%
or, upon rearrangement,%
\begin{equation}
g\left( z\right) =g\left( 0\right) \left( 1\pm g^{-\frac{\alpha -n}{2\alpha }%
}\left( 0\right) \frac{\alpha -n}{2\alpha }\sqrt{\frac{2\alpha }{\alpha +n}}%
z\,\right) ^{\frac{2\alpha }{\alpha -n}}\,.
\end{equation}%
The boundary condition is 
\begin{equation}
{j}_{0}^{\ast }=-\left. \frac{dg}{dz}\right\vert _{z=0}=\mp \sqrt{\frac{%
2\,\alpha }{\alpha +n}}\;g^{\frac{\alpha +n}{2\alpha }}\left( 0\right) .
\end{equation}%
Since we are interested in the circumstance that the injection rate is
positive we must take the lower sign so%
\begin{equation}
g\left( z\right) =\left( {j}_{0}^{\ast }\sqrt{\frac{\alpha +n}{2\alpha }}%
\right) ^{\frac{2\alpha }{\alpha +n}}\left( 1-\frac{\alpha -n}{2}\frac{z}{%
z_{0}}\,\right) ^{\frac{2\alpha }{\alpha -n}},\;\;z_{0}=\alpha {j}_{0}^{\ast 
\frac{\alpha -n}{\alpha +n}}\left( \frac{\alpha +n}{2\alpha }\right) ^{\frac{%
\alpha }{n+\alpha }}\,,
\end{equation}%
or, rewriting the result in terms of the physical variables, 
\begin{equation}
c\left( r\right) =\left( {j}_{0}^{\ast }\sqrt{\frac{\alpha +n}{2\alpha }}%
\right) ^{\frac{2}{\alpha +n}}\left( 1-\frac{\alpha -n}{2}\,\frac{r}{r_{0}}%
\right) ^{\frac{2}{\alpha -n}},\;\;r_{0}=\alpha {j}_{0}^{\ast \frac{\alpha -n%
}{\alpha +n}}\left( \frac{\alpha +n}{2\alpha }\right) ^{\frac{\alpha }{%
n+\alpha }}\sqrt{\frac{\,D}{k}}\,.  \label{physical}
\end{equation}%
There are two cases that must be distinguished depending on whether $%
n>\alpha $ or $\alpha >n$. In the first case the solution has \emph{infinite
support} and is a simple algebraic decay%
\begin{equation}
c\left( r\right) =\left( {j}_{0}^{\ast }\sqrt{\frac{\alpha +n}{2\alpha }}%
\right) ^{\frac{2}{\alpha +n}}\left( 1+\frac{n-\alpha }{2}\,\frac{r}{r_{0}}%
\right) ^{-\frac{2}{n-\alpha }},\;\;n>\alpha \,.  \label{infinite}
\end{equation}%
The second, more complicated case occurs when $\alpha >n$. Then it is clear
from Eq.(\ref{physical}) that the concentration will, in general become
imaginary and in all cases its magnitude will increase without bound for
sufficiently large $r$. The only way to avoid this unphysical behavior is if
the solution has \emph{finite support} so that%
\begin{equation}
c\left( r\right) =\left( {j}_{0}^{\ast }\sqrt{\frac{\alpha +n}{2\alpha }}%
\right) ^{\frac{2}{\alpha +n}}\left( 1-\frac{\alpha -n}{2}\,\frac{r}{r_{0}}%
\right) ^{\frac{2}{\alpha -n}}\Theta \left( \frac{2}{\alpha -n}%
r_{0}-r\right) ,\;\;\alpha >n\,,  \label{finite}
\end{equation}%
where the step function $\Theta \left( x\right) =1$ for $x>0$ and zero
otherwise. Noting that%
\begin{eqnarray}
\frac{d}{dx}f\left( x\right) \Theta \left( x\right)  &=&f^{\prime }\left(
x\right) \Theta \left( x\right) +f\left( 0\right) \delta \left( x\right) \,,
\notag \\
\frac{d^{2}}{dx^{2}}f\left( x\right) \Theta \left( x\right)  &=&f^{\prime
\prime }\left( x\right) \Theta \left( x\right) +f^{\prime }\left( 0\right)
\delta \left( x\right) +f\left( 0\right) \delta ^{\prime }\left( x\right) \,,
\notag
\end{eqnarray}%
it is clear that  \Ex{finite} can only be an acceptable solution to the 
 steady state equation \Ex{ss_eq} if the first two derivatives of the coefficient
of the step function vanish at  $r= \frac{2 \,r_{0}}{\alpha - n}$. This simply imposes the
requirement on the exponent that $\frac{2\alpha }{\alpha -n}-2>0$ which is
always true provided that $n>0$, as was already required. Thus, the final,
physically valid solution with finite support {\Ex{finite}} is restricted to a 
range of values of the coefficients  $\alpha > n > 0$.

We note that the solutions { \Ex{infinite} and \Ex{finite}} 
can be expressed as $q$-exponentials, \newline
$e_{q}\left( x\right) =\left( 1+\left( 1-q\right) x\right) ^{\frac{1}{1-q}%
}~\Theta \left( 1+\left( 1-q\right) x\right) $ with the identification $q=%
\frac{n-\alpha }{2}+1$ and that  $q>1$ gives the
case of infinite support while  $1>q$ gives the case of finite support. From the
properties of the $q$-exponential we know that for $q=1$ the decay of the
concentration will be expontial, $c\left( r\right) =\left( {j}%
_{0}^{\ast }\right) ^{\frac{1}{\alpha }}\,e^{-r/r_{0}}$ with $r_{0}=\alpha 
\sqrt{\frac{\,D}{k}}$. This of course includes the steady state solution of the classical
reaction-diffusion equation with $\alpha =n=1$.

\begin{figure}[tbp]
{\includegraphics[angle=-90,scale=0.3]{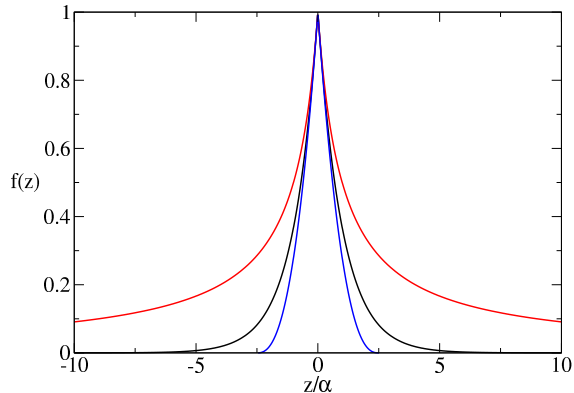}}

\caption{Steady state: $c(r) / c(0) = \exp \left(- {r /r_0}\right)$ ($q=1$; black);
$c(r) / c(0) = \left( 1 + \frac{n-\alpha}{2} \frac{r}{r_0}  \right)^{-\frac{2}{n-\alpha}}$ 
for $n = \alpha + 2$ (infinite support, $q=2$; red) and $\alpha = n + 0.8$ 
(finite support, $q= 0.6$; blue).} \label{fig0}
\end{figure}

The physical interpretation of these results can be understood as follows:
increasing $n$ decreases the extinction rate (since the reaction term goes
like $c^{n}$ and $c<1$) while increasing $\alpha $ decreases the rate of
diffusion {\ (this is easily seen from the scaling $r\sim t^{\gamma /2}$ or
by writing the diffusion term as $\frac{\partial }{\partial r}D\frac{%
\partial c^{\alpha }}{\partial r}=\frac{\partial }{\partial r}\left( \alpha
\,Dc^{\alpha -1}\right) \frac{\partial c}{\partial r}$, so the effective
diffusion coefficient goes like $c^{\alpha -1}$)}. Hence, making $n$ large
or $\alpha $ small leads to infinite support: diffusion is fast, extinction
is slow. The converse, making $n$ small or $\alpha $ large leads to finite
support because diffusion is slow and extinction is fast. The resulting
steady state profiles are compared in\ Fig. \ref{fig0}.

\subsection{Robustness of the steady state}

\label{robustness}

The question of robustness is an important issue as discussed by Eldar et
al. {\cite{eldar}} and by Yuste et al. {\cite{yuste}} in particular for
morphogen gradient formation as precursor to cell differentiation.
Robustness is a measure of the strength of the steady state profile versus
changes in the variables controlling input flux and degradation, such as $%
j_{0}$ and $k$. The cited authors characterized it as the quantity ${%
\mathcal{R}}_{b}=d\,\left\vert \partial L/\partial \log b\right\vert ^{-1}$
where $d$ is a characteristic microscopic length (e.g. the cell size) and $b$
denotes $j_{0}$ or $k$; $L$ is the distance at which the steady state $c(r)$
takes a given value and is obtained by inversion of the steady state
solution $c(r)_{r=L}$. A high value of ${\mathcal{R}}_{b}$ is an indication
of the buffering capacity against changes in the input flux and degradation
rate. Here, however, we prefer to consider directly the relative change in
the concentration at point $r$ due to a change in the value of quantity $b$,
thereby defining the (position-dependent) \textit{sensitivity} to parameter $b$ as 
\begin{equation}
{\mathcal{S}}_{b}\left( r\right) =\frac{\partial \log c\left( r\right) }{%
\partial \log b}\,.  \label{sensitivity}
\end{equation}

For $n>\alpha $, the case of infinite support, a short calculation gives the
sensitivity as%
\begin{equation}
{\mathcal{S}}_{j_{0}}\left( r\right) =\frac{2}{\alpha +n}\;\frac{1}{1+\frac{%
n-\alpha }{2}\,\frac{r}{r_{0}}}\;\;\;;\;\;\;n\geq \alpha ,
\end{equation}%
and for $n=\alpha =1$, i.e in the classical case of exponential decay, this
becomes 
\begin{equation}
{\mathcal{S}}_{j_{0}}\left( r\right) =1\;\;\;;\;\;\;n=\alpha =1,
\end{equation}%
which we will take as a reference point. One also gets exponential decay for
the more general condition $n=\alpha $ (see section \ref{ss_sol}), but in
this case we find%
\begin{equation}
{\mathcal{S}}_{j_{0}}\left( r\right) =\frac{1}{n}\;\;\;;\;\;\;n=\alpha ,
\end{equation}%
so that even though the decay is exponential, it is nevertheless true that
increasing the nonlinearity of the process decreases the sensitivity of the
concentration to variations in the injection rate. Note that the general
result for infinite support is bounded by 
\begin{equation}
{\mathcal{S}}_{j_{0}}\left( r\right) \leq \frac{2}{\alpha +n}%
\;\;\;;\;\;\;n\geq \alpha ,
\end{equation}%
so that - independent of position - increasing nonlinearity in either the
diffusion process or in the extinction process has the effect of buffering
the concentration against changes in the rate at which material is injected.

The case of finite support, $\alpha <n$, is more complicated. A simple
calculation gives%
\begin{equation}
{\mathcal{S}}_{j_{0}}\left( r\right) =\frac{1}{1-\frac{\alpha -n}{2}\;\frac{r%
}{r_{0}}}\frac{2}{\alpha +n}\;\;\;;\;\;\;\alpha >n,  \label{S_a>n}
\end{equation}%
so that there are two effects at work:\ decreasing sensitivity with
increasing nonlinearity, as above, and increasing sensitivity with
increasing distance from the source. In fact, in this case we find 
\begin{equation}
{\mathcal{S}}_{j_{0}}\left( r\right) >1\Longleftrightarrow r>r_{\ast }\equiv 
\frac{2}{\alpha -n}\left( 1-\frac{2}{\alpha +n}\right) r_{0}.
\end{equation}%
Clearly, this is only relevant if the right hand side is less than $r_{0}$.
For $n<1$, this is always the case:\ i.e., there is always a region of
enhanced sensitivity in the range $r_{\ast }<r<r_{0}$. For $n>1$, there is a
region of enhanced sensitivity for  
\begin{equation}
\alpha >\alpha _{\ast }\equiv 1+\sqrt{\left( n-1\right) \left( n+3\right) }%
=\allowbreak n+2-\frac{2}{n}+...\;\;\;;\;\;\;\alpha >n>1.
\end{equation}%
Only for the restricted range $\alpha _{\ast }>\alpha >n>1$ is there no
region of enhanced sensitivity for the case of finite support. 

In summary, we find that (i) for infinite support, $n\geq \alpha $,
increasing nonlinearity \emph{always} decreases sensitivity of the
concentration to the injection rate; (ii) the same holds true for the case
of finite support when $\alpha _{\ast }>\alpha >n>1$; (iii) the case of
finite support will, for $n<1$ or $\alpha >\alpha _{\ast }$ show enhanced
sensitivity in the region  $r_{\ast }<r<r_{0}$ . 

\section{Comparison to simulation and experiment}

\subsection{Numerical solution of master equation}

\begin{figure}[tbp]
{\includegraphics[angle=-90,scale=0.25]{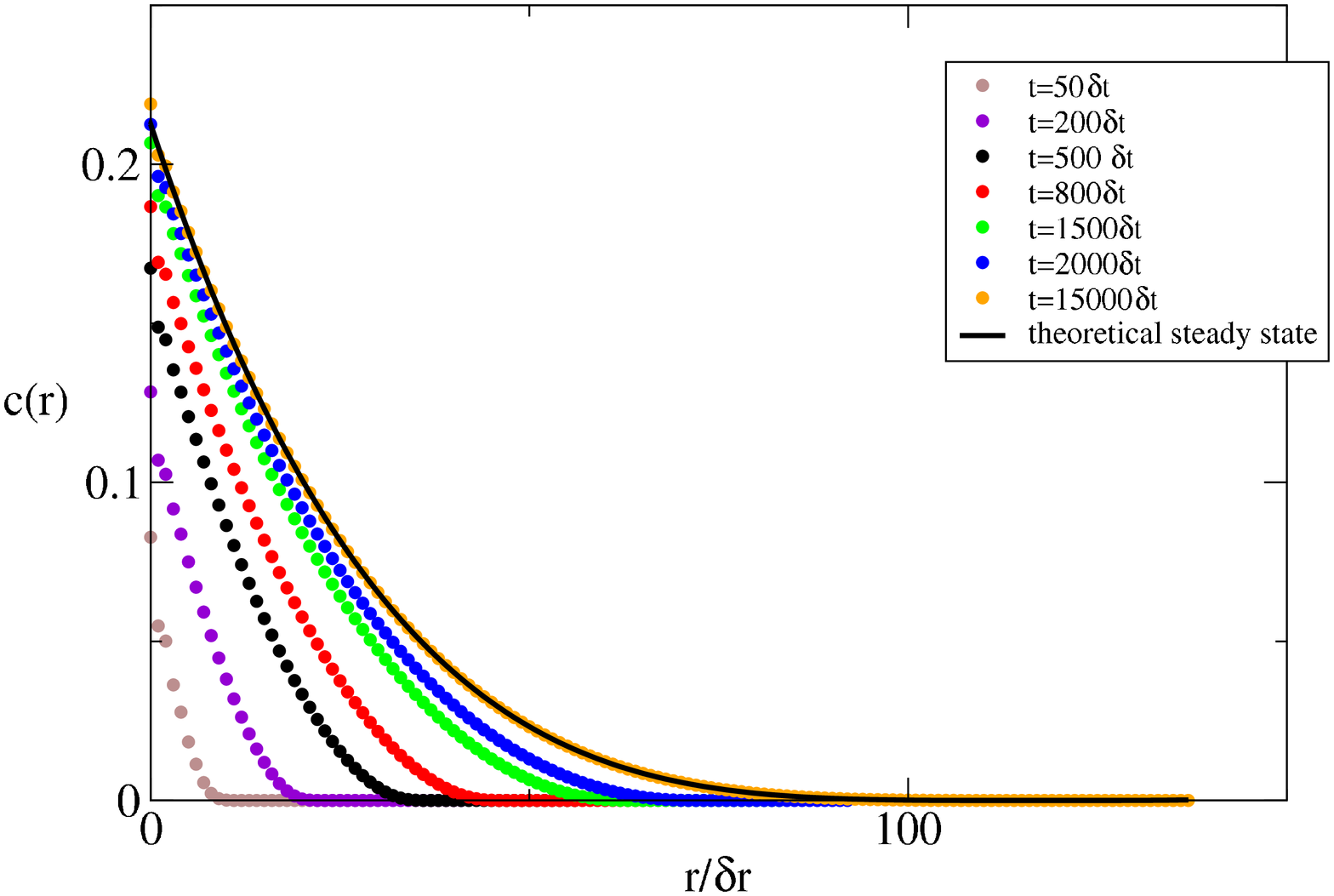}} 
{\includegraphics[angle=-90,scale=0.25]{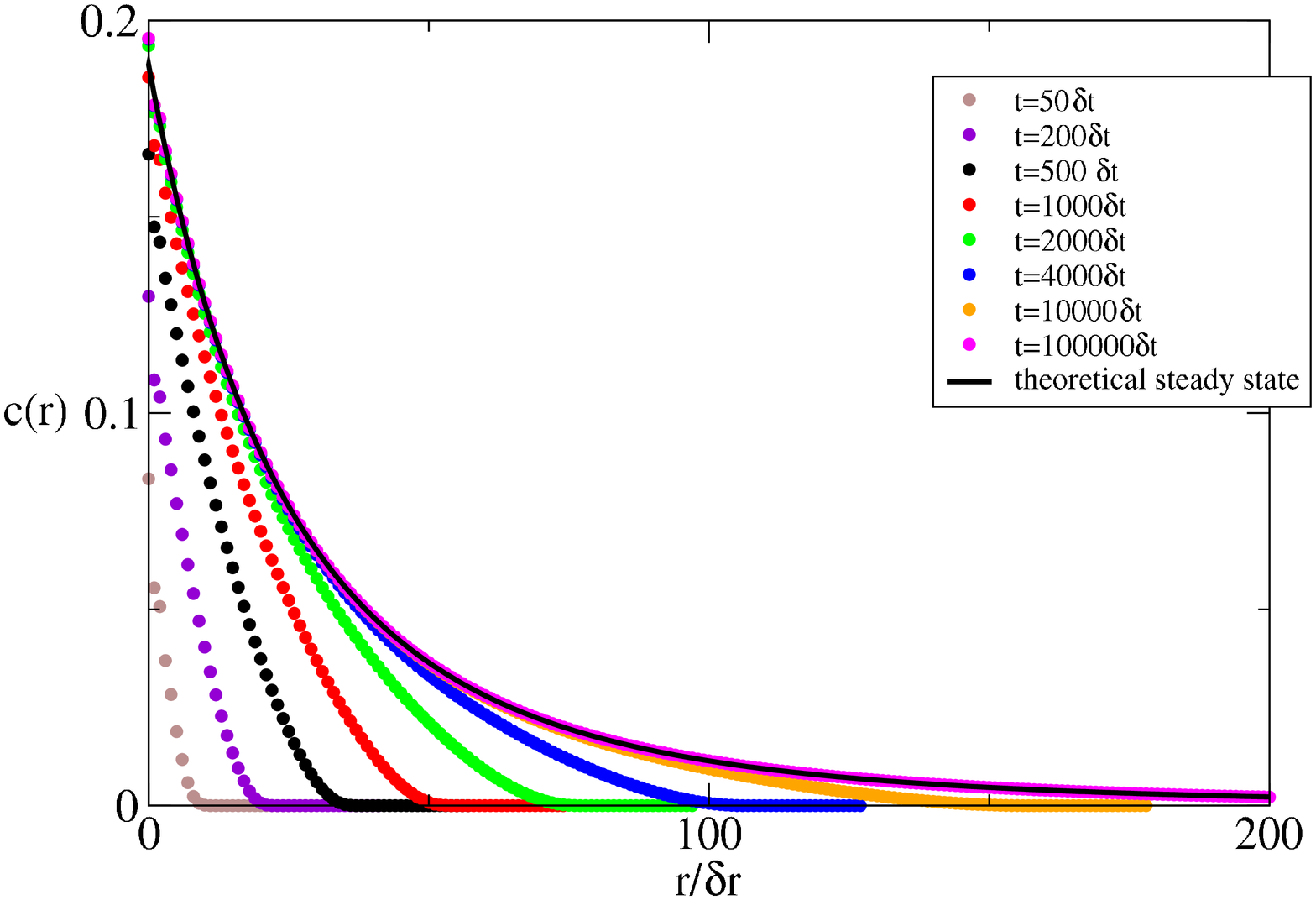}} 

\caption{Numerical and analytical solutions for the steady state profile.
Left panel:  Case of finite support. Numerical solution of the Master Equation \Ex{Master_Eq}
where $p_j\,F= p_j\,c_j^{\alpha -1}$ with $p_j = 0.2$ for $ j\in[-2,+2] $ and   $\alpha = 1.5$  and $G=p_R\,c^{n}$ with $n=1$ and $p_R = 10^{-3}$, for 
$t= 50, 200, 500, 800, 1.5  \times 10^3, 2  \times 10^3, 1.5  \times 10^4$ time steps (symbols);
the boundary condition is finite flux at $r=0$ and the initial condition is zero concentration everywhere. 
 Analytical steady state solution \Ex{finite} (black curve).  
Right panel :  Case of infinite support. Same as left panel except $n = 2$ and $p_R = 10^{-2}$, for
$t= 50, 200, 500, 10^3, 2 \times 10^3, 4 \times 10^3, 10^4,1 \times 10^5$ time steps
(symbols) and steady state solution \Ex{infinite} (black curve). 
Note that there are no adjustable parameters in either case.} \label{fig2}
\end{figure}

\begin{figure}[tbp]
{\includegraphics[angle=-90,scale=0.25]{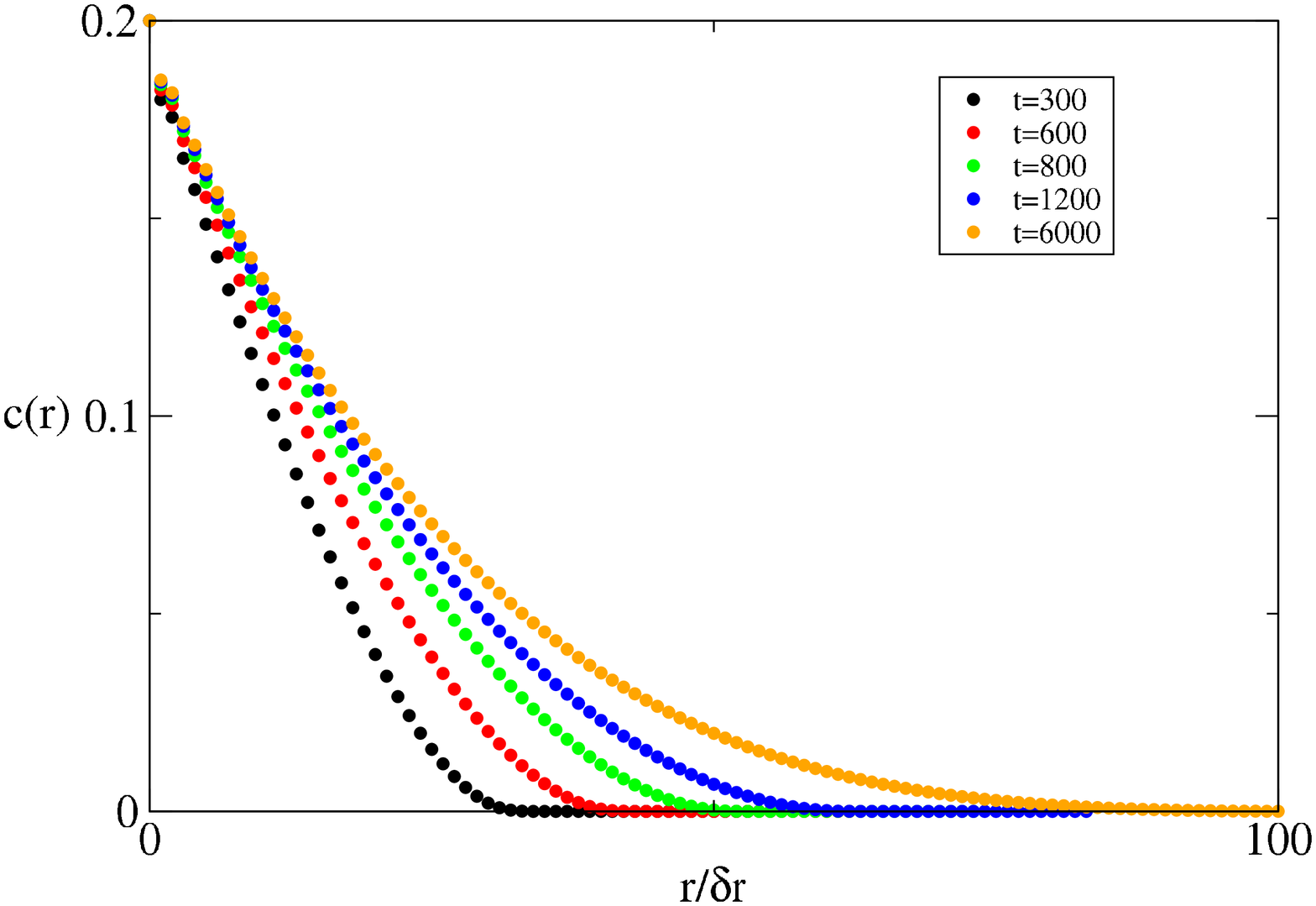}} {%
\includegraphics[angle=-90,scale=0.25]{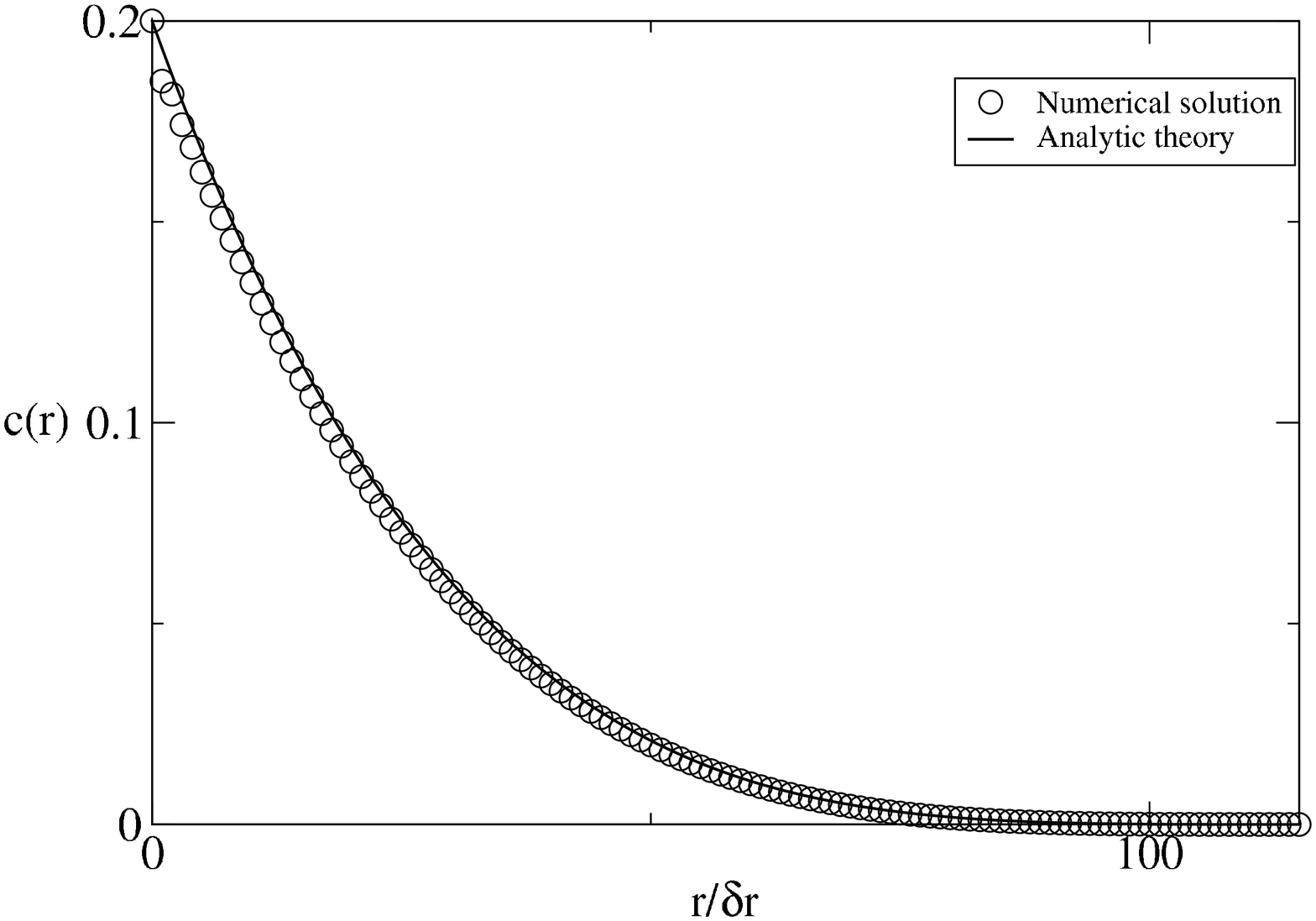}} 

\caption{Numerical and analytical solutions for finite support profile ($n < \alpha$)  
with boundary condition of fixed $c(0)$.
Left panel: Numerical solution of the Master Equation \Ex{Master_Eq}
where $p_j\,F= p_j\,c_j^{\alpha -1}$ with $p_j = 0.2$ for $ j=[-2,+2] $ and   $\alpha = 1.5$  
and $G=p_R\,c^{n}$ with $n=1$ and $p_R = 10^{-3}$, for 
$t= 3\times 10^2, 6\times 10^2, 8\times 10^2, 1.2\times 10^3, 6\times 10^3$ time steps. 
Right panel: Comparison between numerical solution of the Master Equation \Ex{Master_Eq} 
for $t=6\times 10^3$  time steps (open circles) and analytical steady state 
solution \Ex{finite} (black curve).  Note that there are no adjustable parameters.}
\label{fig1}
\end{figure}

We have performed numerical computation of the master equation (\ref%
{Master_Eq}) in order to verify three aspects of this theory: first, that
the non-linear dynamics eventually leads to a steady state; second, that the
steady state is independent of the initial conditions and third, that our
analytic, continuum result is a good representation of the steady state.
Figure \ref{fig2} shows the result of solving the master equation with an
initial condition $c(r)=0$ and with constant flux at the origin for two
cases: one with finite support, $n<\alpha $, and one with infinite support, $%
n>\alpha $. In both cases, we do indeed find that at long times the system
settles into a steady state that is well-described by the analytic results, (%
\ref{infinite}) and (\ref{finite}). Note that, in the case of infinite
support, one must go to somewhat longer times to reach the steady state. To
test that the sensitivity of the steady state to the boundary conditions,
the calculations were repeated with a boundary condition of fixed value of
the concentration at $r=0$. The result for the case of finite support is
shown in Fig. \ref{fig1} where it is again seen that the system reaches a
steady state and that the steady state is that of the continuum theory.
Similar results were found for the case of infinite support. This comparison
of numerical and analytical results therefore shows good agreement between
the continuum approximation and the discrete microscopic dynamics and
furthermore provides evidence that the steady state is unique.

\subsection{Comparison to experiment}

\begin{figure}[tbp]
{\includegraphics[angle=-90,scale=0.3]{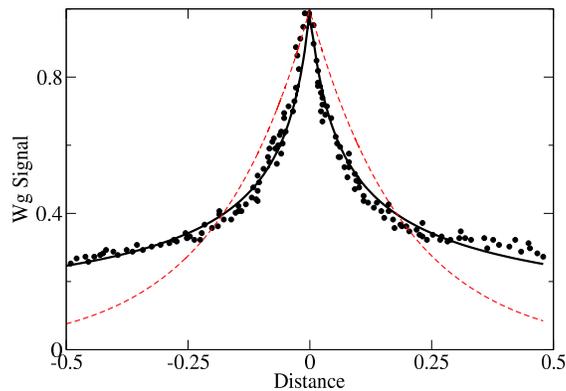}} 

\caption{Experimental data (black dots) from Han {\it et al}, Fig.6.A  in \cite{han} 
 from the fluorescence intensity of the Wg protein (vertical axis; normalized values) 
versus distance (in a.u.; horizontal axis) measured from the anterior-posterior axis along 
the dorsoventral direction in the posterior compartment of the {\it Drosophila} wild-type 
wing disc \cite{han}. The black curve is the
best-fit of the theoretical steady state \Ex{infinite} with $ n - \alpha \simeq 3.8 $.
For comparison the dashed curve shows the best-fit exponential profile.} %
\label{fig3}
\end{figure}

\begin{figure}[tbp]
{\includegraphics[angle=-90,scale=0.25]{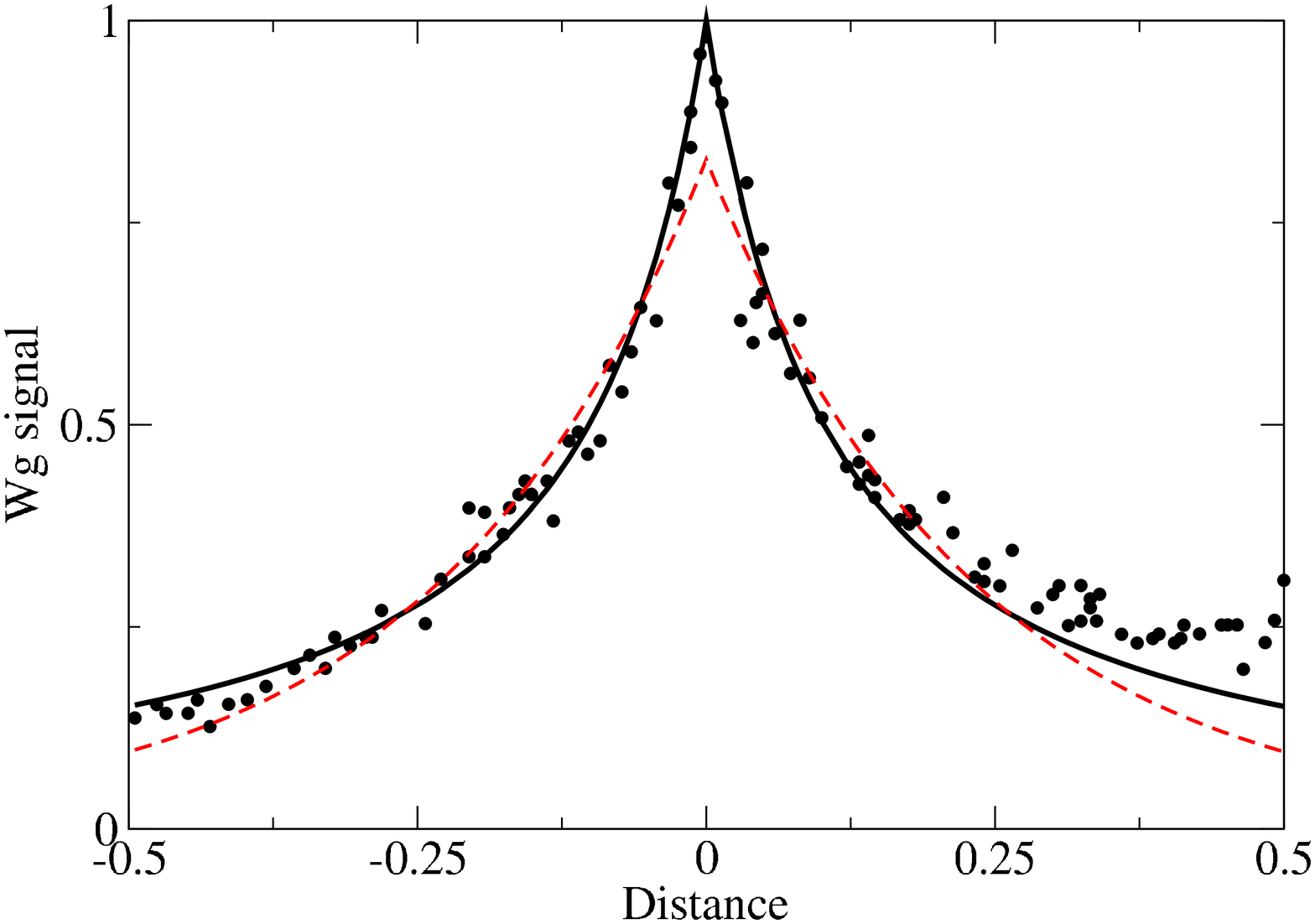}} 
{\includegraphics[angle=-90,scale=0.25]{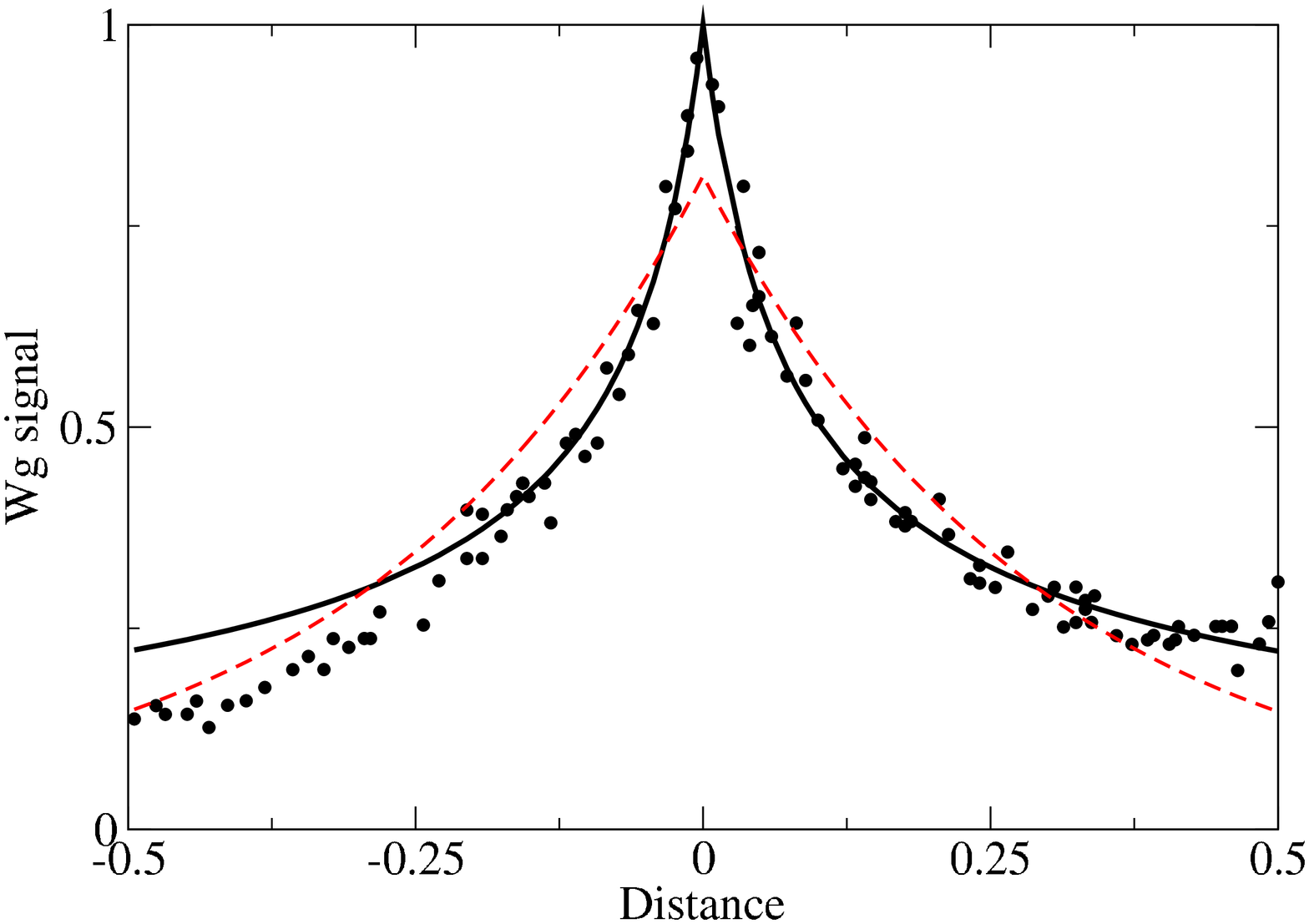}} 

\caption{  Same as Fig.\ref{fig3} for experimental data (black dots) from Han 
{\it et al}, Fig.6.B in \cite{han} for a mutant strain. Fit of theoretical steady state 
\Ex{infinite} (black curve) to the  experimental data;  because of the obvious 
asymmetry of the data along the dorsoventral axis, the left panel shows a fit 
based only on the data for negative distances, giving $n - \alpha \simeq 1.7 $,  
and the right panel shows a fit to data for positive distances, giving 
$ n - \alpha \simeq 3.3 $. In both cases, a best-fit to an exponential decay is shown as the
dashed curves.} \label{fig4}
\end{figure}

As an application of the theory we compare our analytical solution for the
steady state with experimental results obtained from measurements performed
in the \textit{Drosophila} wing disc where morphogens are produced by a
subset of cells wherefrom they diffuse and are degraded thereby forming a
concentration gradient whose profile shape appears crucial for subsequent
cell specification {\cite{eldar}}. This situation corresponds to the
reaction-diffusion theory presented in the present article. Experimental
results given in {\cite{han}} present the intensity signal  of the Wg morphogen as 
a function of distance from the source obtained by image processing showing the profile of
the diffusing protein in selected regions of the \textit{Drosophila} wing
disc. In the absence of numerical data, we processed the signal images to
obtain the data shown in Figs. \ref{fig3} and \ref{fig4} where they are
compared to our analytical results. Clearly we find that the sub-diffusive
nonlinear reactive steady state profile (\ref{infinite}) with infinite
support reproduces very well the experimental data indicating slow
degradation combined with extended sub-diffusion. In all cases, we also show
best-fits to an exponential of the form $f(z)=Ae^{(-B|z|)}$ and it is clear
that the experimental data are very poorly fit by an exponential decay. 

\section{Comments}

\label{comments}

We derived the nonlinear reaction-diffusion equation starting from
Einstein's microscopic model where the diffusing particles are also subject
to an annihilation reactive process. The nonlinear reaction-diffusion
equation was obtained under the demand that scaling be satisfied for
diffusive motion wherefrom a relation follows between the scaling exponent
and the nonlinear exponents whose range of possible values exhibit the
signature of sub-diffusion. While full scaling should in principle be
satisfied for the space-time dependent equation, this requirement can be
relaxed between the reaction term exponent and the scaling exponent for the
steady state equation. This observation is important for the analysis of the
R-D steady state solutions which take the form of a power law with in one
case infinite support and in the other case finite support.

We discussed the sensitivity of the steady state versus changes in the input
flux and we found that profiles with infinite support show minimal
sensitivity, and such profiles with infinite support were shown to
correspond to experimental observations. On the other hand we showed that
profiles with finite support should exhibit stronger sensitivity to input
flux changes, and it seems that such profiles with finite support have not
been observed in morphogen gradient formation. This observation may suggest
that extreme sensitivity excludes this type of profile in natural morphogen
gradient formation because degradation is too fast with respect to diffusion
in order to establish the necessary gradient for subsequent cell
differentiation.

\acknowledgements {This work was supported in part by the European Space
Agency under contract number~ESA AO-2004-070.}

\bigskip


\end{document}